\documentclass[9pt,twocolumn,twoside]{opticajnl}
\journal{opticajournal} 

\setboolean{shortarticle}{true}

\usepackage{lineno}

\title{Rubber band filters: optimal padding without edge artifacts}

\author[1]{Zhenyang Xiao}
\author[1,*]{David Burghoff}

\affil[1]{Cockrell School of Engineering, The University of Texas at Austin, Austin, Texas, 78712}

\affil[*]{burghoff@utexas.edu}

\begin{abstract}
Bandpass filtering techniques are widely used in spectroscopy. However, conventional symmetric-padding filtering methods introduce boundary artifacts that distort the signal at the edges. We present a rubber band filter: a robust method for achieving band‐limited filtering without these detrimental edge artifacts. The technique applies an optimal padding scheme during the filtering process, thereby overcoming longstanding challenges in achieving artifact‐free filtering. Importantly, it is iterative and requires only a few extra Fourier transforms over conventional approaches. We demonstrate its superiority and versatility by applying it to three spectroscopic examples---time‐domain spectroscopy, Fourier‐transform spectroscopy, and dual‐comb spectroscopy. 
\end{abstract}

\setboolean{displaycopyright}{false} 

\begin{document}

\maketitle

\begin{figure}[ht]
\centering
\includegraphics{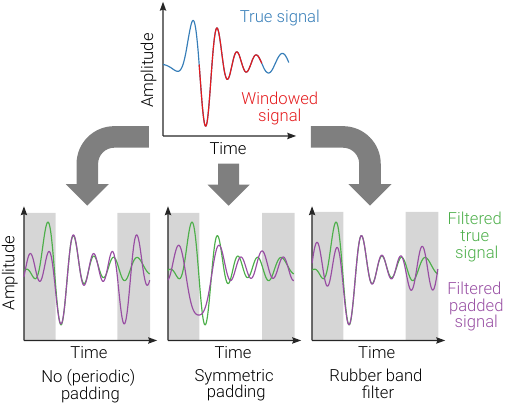}
\caption{Comparison between low-pass filtering without padding (left), with symmetric padding (middle), and with rubber band filtering (right). The shaded region shows the region where the signal is padded. The true signal is the signal without any padding.}
\label{figcompare}
\end{figure}

Bandpass filtering techniques are widely used in imaging \cite{lu2014medical,fauvel2012advances,manolakis2003hyperspectral,chandel2013image} and spectroscopy \cite{burghoff2016computational,nazarian2022real,li2014review,lasch2012spectral}. By applying an ideal filter to a spectroscopic signal, noise or unwanted spectral information can be removed. However, if the signal is not padded prior to analysis, this introduces artifacts due to the discontinuity between the beginning and end of the signal, as no padding is equivalent to periodic padding. As shown in Fig.~\ref{figcompare}, such discontinuities lead to distortions like spectral leakage and baseline shifts, especially when the signal has a sharp slope at the window edges. Apodization (window) functions are typically used to mitigate these effects, but they are not suitable for all data types. For example, when filtering a signal like the phase of a dual-comb spectroscopy signal \cite{coddington2016dual,burghoff2019generalized}, apodization is inappropriate since the phase of a sinusoid is inherently non-zero and should have a slope. To mitigate these effects, symmetric padding is often used, in which the signal is flipped and extended. It is an easy-to-implement method that generally performs better than non-padding or periodic padding techniques \cite{fu2002low,dzhezyan2021symmetrical,bovik2009basic,chakrabarti1999dwt}. However, symmetric padding also introduces issues when the slope is steep at the boundary between the signal and the padded region. These edge artifacts are evident in the differences between the true filtered signal and the symmetrically padded version. Such distortions can undermine the reliability of spectral analysis and degrade the performance of downstream applications, like phase correction.

To address these challenges, a rubber band filter is introduced---a signal filtering method designed to achieve a band-limited signal while eliminating edge artifacts by using \textit{optimal} padding. An example is shown in Fig.~\ref{figcompare}, in which a windowed signal is filtered without padding (equivalent to periodic padding), with symmetric padding, and with a rubber band filter. The rubber band filtered signal preserves its integrity far better than with other filtering methods, as evidenced by the much smaller difference between the filtered true signal and the filtered signal. By extending the signal through symmetric padding and applying a carefully designed weighting scheme that de-emphasizes the padded regions, especially the edge region, our approach transforms the filter into a weighted least-squares problem in the time domain. Transforming this problem into the frequency domain yields a linear system whose solution provides the filtered Fourier amplitudes. This method not only preserves the spectral integrity but also effectively removing the redundancy information, thereby offering a robust tool for accessing optical measurement signals within a desired frequency range.
\section{Method}

The core of our approach, as illustrated in Algorithm~\ref{alg:rubberband_theory}, is to transform the filtering problem into a fitting problem, one constrained by the bandpass filter. We extend the signal beyond the valid region (typically with symmetric padding) but \textit{de-weight} the padded region in the fit. In this way, the signal is allowed to vary however it needs to in the padded region to best fit the original data while remaining bandlimited, but is not tightly constrained outside the valid region. If we call the padded signal \(s(t_n)\), the weighting function \(w(t_n)\), and the filtered signal \(s_f(t_n)\), then cost function $J$ is taken as a standard norm and the filtered signal can be expressed as a discrete Fourier transform:
\begin{equation}\label{eq:Vf}
s_f(t_n) = \frac{1}{N}\sum_{\omega_k\in \Omega} F(\omega_k)\, e^{\,i\,\omega_k\,t_n}
\end{equation}
\begin{equation}\label{eq:cost}
J = \sum_{n=1}^{N} w(t_n)\,\left| s_f(t_n)-s(t_n)\right|^2.
\end{equation}
In the above expression, \(\Omega\) is used to indicate the set of frequencies in the passband and \(F(\omega_k)\) are the unknown Fourier amplitudes. Our goal is to select the set of $F(\omega_k)$ that minimize the cost function. The only input to this model, aside from the choice of filter, is $w(t_n)$, which is typically set to 1 in the window, and a small number outside the window.

\begin{algorithm}
\caption{Rubber-Band Filter}\label{alg:rubberband_theory}
\begin{algorithmic}[1]
  \State Set $s(t_n)$ as symmetrically-padded $s_{in}(t_n)$, $w(t_n)$ as 1 inside the window and $w(t_n)\ll1$ outside the window
  \State Compute $S(\omega_k)=\mathcal{F}\{\,s\}$,   $W(\omega_k)=\mathcal{F}\{\,w\}$, 

  \State Solve the linear system $A F = b \Leftrightarrow
W*F=W*S$ using a conjugate gradient method (computing $Ax=W*x$ using FFTs):
    \begin{itemize}
        \item[] Initialize: $r_0 = b - A\,F_0$, \quad $p_0 = r_0$
        \item[] \textbf{for} $m = 0, 1, \ldots$ until convergence \textbf{do}
        \begin{itemize}
            \item[] $\alpha_m = \frac{r_m^\top r_m}{p_m^\top A\,p_m}$
            \item[] $F_{m+1} = F_m + \alpha_m\, p_m$
            \item[] $r_{m+1} = r_m - \alpha_m\, A\,p_m$
            \item[] \textbf{if} $\|r_{m+1}\|$ is sufficiently small \textbf{then break}
            \item[] $\beta_m = \frac{r_{m+1}^\top r_{m+1}}{r_m^\top r_m}$
            \item[] $p_{m+1} = r_{m+1} + \beta_m\, p_m$
        \end{itemize}
        \item[] \textbf{end for}
    \end{itemize}

  \State $s_f(t)\leftarrow\mathcal{F}^{-1}\{F(\omega_k)\}$                  
\end{algorithmic}
\end{algorithm}

Because the cost function $J$ is quadratic and convex, it can be minimized efficiently with global convergence guaranteed. In fact, we will show that only a handful of fast Fourier transforms are needed to perform the inversion. To see why, we substitute the discrete transform of \eqref{eq:Vf} into the cost
function \eqref{eq:cost} and take the stationary condition  
\(\frac{\partial J}{\partial F^*(\omega_i)}  = 0\) for each index
\(\omega_i\in\Omega\), yielding:
\begin{equation}\label{eq:conv}
\sum_{\omega_k} W \left( \omega_i-\omega_k \right) F(\omega_k)
= \sum_{\omega_k} W \left( \omega_i-\omega_k \right) S(\omega_k)
\quad \forall\, \omega_i \in \Omega.
\end{equation}
where \(S(\omega_k)\) is the discrete Fourier transform of the original padded signal \(s(t_n)\), and \(W(\omega_k)\) is the Fourier transform of the weight function. The summation over $\omega_k$ is taken over all frequencies in the Nyquist band: while $F(\omega_k)=0$ for $\omega_k \notin \Omega$,  \(S(\omega_k)\) and  \(W(\omega_k)\) are not band-limited.

Both sides of the stationarity condition represent a convolution in frequency, and so they can be compactly written as $W * F = W * S$. This result can also be expressed in matrix form as
$A F = b$, where the matrix \(A\) and vector $b$ have elements
\begin{align}
A_{ij} &= W\left(\omega_i - \omega_j \right), \label{A} \quad \forall\, \omega_i,~\omega_j\in \Omega.\\
b_i &= \sum_{\omega_k} W\left( \omega_i-\omega_k \right) \, S(\omega_k), \quad \forall\, \omega_i \in \Omega.\label{B}
\end{align}
Of course, at this point the optimal $F$ could be solved using a standard exact method for computing $A^{-1} b$. However, for most signals of interest, this is impractical, as the matrix $A$ scales with $\mathcal{O}(N^2)$ in the signal size N. Instead, a conjugate gradient approach is employed \cite{nazareth2009conjugate}. While we do not elaborate on the details here for the sake of brevity, the most important aspect is that it is iterative and only requires forward calculations of $AF$ to compute $A^{-1} b$. Since $AF$ represents the convolution $W * F$, it can be evaluated extremely efficiently using Fourier transforms. Adequate convergence is typically achieved after a handful of iterations, so the final performance retains the $\mathcal{O}(N \log N)$ performance of fast Fourier transforms. Once the optimal Fourier amplitudes \(F(k)\) are determined, the filtered signal is reconstructed by applying the inverse fast Fourier transform, yielding a band-limited signal \(s_f(t)\) that is free from the edge artifacts typically introduced by conventional symmetric padding methods.

\section{Results}

\begin{figure}[ht]
\centering
\includegraphics{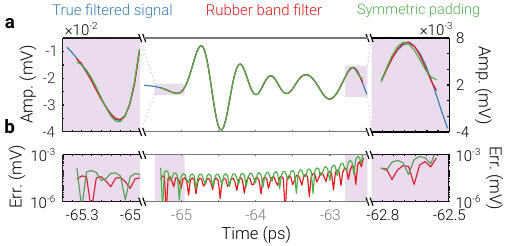}
\caption{\textbf{a}. Comparison of low-pass signal filtering applied to a terahertz time-domain signal extracted from a complete measurement. The true filtered signal was generated by low-pass filtering the full, unpadded signal, serving as an ideal reference. The center normalized signal is unit-less, the shaded region shows a zoomed-in view of the signal at the edges.
\textbf{b}. Comparison of the deviations from this true filtered signal using different filtering methods.}
\label{figTDS}
\end{figure}

To evaluate the performance of the rubber band filter, we use it to filter several spectroscopic signals \cite{xiao2022optical,xiao2025delay,roy2024fundamental,burghoff2019generalized} with unwanted spectral information. First, as illustrated in Fig.~\ref{figTDS}, we demonstrate it on a terahertz time-domain spectroscopy signal that spans up to 4.5\,THz, with frequencies above 4.5\,THz considered as noise. The true filtered signal is defined as the original signal before windowing that has been filtered with the same cut-off frequencies—this serves as the ideal reference for comparison. To illustrate the difference between conventional symmetric padding and our proposed approach, the signal was truncated near its strongest derivative.
For the padded regions, the weight is defined as $w_{\text{pad}}(n) \;=\; 0.03 \, w_{\text{Tukey}}\bigl(n,\,\alpha=0.2\bigr)$, where \(w_{\text{Tukey}}(n,\,\alpha)\) is the standard Tukey window function. \(n\) is the number of padded signal elements. By multiplying the Tukey window by a small constant \(0.03\), the discontinuity at the junction in the padded region is further de-emphasized, thus minimizing artifacts introduced at the junctions between the original data and the padding. The windowed data is assigned a weight of 1.

Figure~\ref{figTDS}a displays the filtered signals. The rubber band filter (red curve) effectively preserves the sharp transitions at the signal end, in contrast to the conventional symmetric padding method, which shows significant distortions in that region. These distortions cause considerable deviations from the true filtered signal. Figure~\ref{figTDS}b further illustrates that the error—defined as the deviation from the true filtered signal—is substantially lower at the boundaries when the rubber band filter is employed.

\begin{figure}[ht]
\centering
\includegraphics{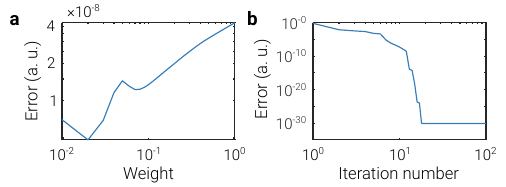}
\caption{\textbf{a}. Normalized error of the filtered signal as a function of the weight parameter, illustrating how increased weighting impacts the overall error.
\textbf{b}. Normalized error plotted against iteration number, showing convergence behavior during the iterative Conjugate Gradient process for solving the band-limited filtering system.
}
\label{figerror}
\end{figure}

In all signal regions, the rubber band filter consistently outperforms the conventional symmetric padding method, yielding errors approximately ten times lower. Although both methods exhibit higher errors at the edges—where sharp discontinuities contribute most to distortion—the adaptive weighting and padding strategy of the rubber band filter markedly reduces these edge artifacts, resulting in a smoother transition and a more faithful reconstruction of the original signal. Moreover, the weight assigned to the padded regions is initially chosen somewhat arbitrarily, but as we will show in Fig.~\ref{figerror}a, minor tuning of these weight parameters can further reduce the overall error relative to the true filtered signal. It essentially takes the role of a regularization constant.
\begin{figure}[ht]
\centering
\includegraphics{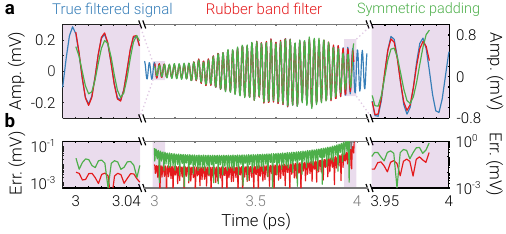}
\caption{\textbf{a}. Comparison of low-pass signal filtering applied to an Fourier Transform Infrared Spectroscopy interferogram signal of a mid-IR quantum cascade laser extracted from a complete measurement. The center normalized signal is unit-less, the shaded region shows a zoomed-in view of the signal at the edges.
\textbf{b}. Comparison of the deviations from this true filtered signal using different filtering methods.}
\label{figFTIR}
\end{figure}

Figure~\ref{figFTIR}a shows the application of our filter to a Fourier-transform infrared spectroscopy signal from a mid-infrared quantum cascade laser spanning approximately 40–45\,THz. Similar performance can be observed in Fig.~\ref{figFTIR}b; however, it is even more evident that the rubber-band filter preserves sharp edge features more accurately than the conventional symmetric-padding method, which introduces significant distortions at the signal boundaries. These distortions arise because the higher frequency of the mid-infrared spectrum (compared to the terahertz range shown in Fig.~\ref{figTDS}) produces a sharper time-domain interferogram, particularly at the edges.

\begin{figure}[ht]
\centering
\includegraphics{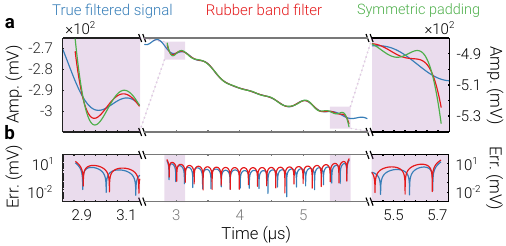}
\caption{\textbf{a}. Comparison of low-pass signal filtering applied to a dual comb spectroscopy data after removing the linear phases. The center normalized signal is unit-less, the shaded region shows a zoomed-in view of the signal at the edges.
\textbf{b}. Comparison of the deviations from this true filtered signal using different filtering methods.
}
\label{figDCS}
\end{figure}

Dual-comb spectroscopy data was also evaluated in Fig.~\ref{figDCS}, where the rubber-band filter was used to filter the phase noise of the dual-comb spectroscopy data. The phase data represent the phase of the offset frequency; the repetition rate shows a very similar result and is therefore not plotted. Figures~\ref{figDCS}a shows the phase comparison after filtering, before and after removing the linear phase for visualization. Figure~\ref{figDCS}b shows the same error information, where the rubber-band filter outperforms by approximately a factor of three. 

To further illustrate the features of our approach, Figure~\ref{figerror} highlights three additional aspects: the effect of the weight parameter on filtering performance, the convergence behavior of the iterative Conjugate Gradient solution for the linear system \(A\,\mathbf{x} = b\), and the padded region comparison between our method and symmetric padding method. In Fig.~\ref{figerror}a, the normalized error is plotted as a function of the weight parameter, demonstrating that an optimal weight effectively balances fidelity to the original signal with the suppression of artifacts at the signal edges. This particular plot evaluates the performance using terahertz time-domain spectroscopy data and applies a Tukey window with \(\alpha = 0.2\); under these conditions, the optimal weight for the constant in front of the Tukey window is found to be 0.06. For different window selections and data sets, the optimal parameters may vary but usually small. 


\section{Discussion}


A key advantage of this approach is in its efficiency. The conjugate gradient method is known to converge very fast, and since the typical rubber band output is not too dissimilar from the naively-filtered spectrum, convergence occurs extremely quickly. Fig.~\ref{figerror}b displays the convergence behavior of the Conjugate Gradient algorithm employed to solve the linear system \(A\,\mathbf{x} = b\). The error is defined as the difference between \(x_k\) and the direct solution \(A^{-1}b\) at each step. As indicated by this figure, it decreases rapidly during the initial iterations, attesting to the efficiency of the Conjugate Gradient method. In the majority of cases we have tested, exact convergence to within the limits of numerical precision is typically achieved within 40 iterations, but it can depend on the data size. For most applications, exact convergence is unneeded, and an adequate level of ~60 dB can be achieved in a handful of iterations.

It is worth mentioning that because the input signal for the rubber band filter process is a padded signal, the output — which is the inverse fast Fourier transform of the optimally bandpass-limited Fourier amplitudes — is also a padded signal in the time domain. However, due to the presence of the weights in the process, the output padded signal differs from the symmetrically padded input, particularly in the junction region. 
With adaptive weighting in the padding region, the padded signal no longer exhibits abrupt changes at the edge; instead, it transitions smoothly with a gently varying derivative. Beyond the padding edge, the filtered trace is almost identical to the original signal, differing only by a small offset. Taken together, these features allow the filtered signal to preserve edge information while avoiding distortion and thus deliver superior filtering performance.

\section{Conclusion}

We have demonstrated a rubber band filter, which can be used to iteratively filter a signal without suffering edge artifacts. These results demonstrate that the rubber band filter not only effectively removed the redundancy signal components but also preserves critical high-fidelity features, particularly at edge regions of abrupt change. This marks a significant improvement over conventional symmetric padding methods in maintaining spectral integrity after filtering the unwanted frequency components.

\section*{Data availability}
Data underlying the results presented in this paper will be published prior to publication, along with example code.
\section*{Acknowledgements}
D.B. acknowledges support from ONR grant N00014-21-1-2735, AFOSR grant no. FA9550-24-1-0349, and NSF grant ECCS-2046772; this research is funded in part by the Gordon and Betty Moore Foundation through Grant GBMF11446 to the University of Texas at Austin to support the work of D.B.
\section*{Competing interests}
The authors declare no competing interests.

\bibliography{ref}



\ifthenelse{\equal{\journalref}{aop}}{%
\section*{Author Biographies}
\begingroup
\setlength\intextsep{0pt}
\begin{minipage}[t][6.3cm][t]{1.0\textwidth} 
  \begin{wrapfigure}{L}{0.25\textwidth}
    \includegraphics[width=0.25\textwidth]{john_smith.eps}
  \end{wrapfigure}
  \noindent
  {\bfseries John Smith} received his BSc (Mathematics) in 2000 from The University of Maryland. His research interests include lasers and optics.
\end{minipage}
\begin{minipage}{1.0\textwidth}
  \begin{wrapfigure}{L}{0.25\textwidth}
    \includegraphics[width=0.25\textwidth]{alice_smith.eps}
  \end{wrapfigure}
  \noindent
  {\bfseries Alice Smith} also received her BSc (Mathematics) in 2000 from The University of Maryland. Her research interests also include lasers and optics.
\end{minipage}
\endgroup
}{}

\end{document}